%% file: main20240816.tex
\documentclass[a4paper, 12pt]{article}
\usepackage{booktabs}
\usepackage[round]{natbib}
\usepackage{array,arydshln}
\usepackage{comment}
\usepackage{xcolor,soul}
\usepackage{here, scalerel}
\usepackage{booktabs}
\graphicspath{ {./images/} }
\usepackage{adjustbox}
\usepackage{multirow}
\usepackage{dcolumn}
\usepackage{lscape}

\usepackage{caption}
\usepackage{subcaption}
\usepackage{tabularx}

\usepackage{hyperref}
\usepackage{amsmath,amsthm,amssymb,setspace,bm}
\usepackage[margin=1.0in]{geometry}

\usepackage{midpage}

\allowdisplaybreaks
\theoremstyle{plain}

\newtheorem{h1}{H1.\ignorespaces}
\newtheorem{h2}{H2.\ignorespaces}

\newtheorem{result}{{\bf  Result}}

\newtheorem*{support}{{\bf  Statistical support}}
	\title{
 Jumping on the bandwagon and off the Titanic: an experimental study of turnout in two-tier voting\footnote{We thank Yuki Yanai and the participants of the Asia Pacific Economic Science Association Meeting in 2022, Asian Meeting of the Econometric Society in 2022, and CREST/LESSAC Workshop in Experimental Economics in 2022 for their helpful comments. Yuki Hamada provided excellent research assistance. 
 Financial support by 
			Investissements d'Avenir, ANR-11-IDEX-0003/Labex Ecodec/ANR-11-LABX-0047, PHC Sakura program, project number 45153XK (JPJSBP 120203208), 
			and
			Joint Usage/Research Center at ISER, Osaka University
			is gratefully acknowledged. This study was approved by the IRB of Osaka University.
		}
	}
	\author{Yoichi~Hizen\thanks{
			Kochi University of Technology. E-mail: \texttt{hizen.yoichi@kochi-tech.ac.jp}.}   
		\and
		Kazuya~Kikuchi\thanks{
			{Tokyo University of Foreign Studies. E-mail: \texttt{kazuya.kikuchi68@gmail.com}}.}
		\and
		Yukio~Koriyama\thanks{
			CREST, Ecole Polytechnique, Institut Polytechnique de Paris. 
			E-mail: \texttt{yukio.koriyama@polytechnique.edu}.}  
		\and
		Takehito~Masuda\thanks{
			Shinshu University. E-mail: \texttt{tmasuda@shinshu-u.ac.jp}.}
	}

	\date{\today}
	
\begin{document}
		
		\maketitle
		
		\begin{abstract}
			We experimentally study voter turnout in two-tier elections when the electorate consists of multiple groups, such as states.
			Votes are aggregated within the groups by the winner-take-all rule or the proportional rule, and the group-level decisions are combined to determine the winner.
			We observe that, compared with the theoretical prediction, turnout is significantly lower in the minority camp (the \textit{Titanic effect}) and significantly higher in the majority camp (the \textit{behavioral bandwagon effect}), and these effects are stronger under the proportional rule than under the winner-take-all rule.
			As a result, the distribution of voter welfare becomes more unequal than theoretically predicted, and this welfare effect is stronger under the proportional rule than under the winner-take-all rule.
		\end{abstract}
		
		\section{Introduction}
		
		Studies of endogenous voter turnout have 
		focused on direct voting in which 
		individuals' 
		votes are aggregated directly to make a 
		social 
		decision. However, there are also cases 
		where 
		social decision-making takes the form of 
		two-tier 
		voting: votes are aggregated separately in 
		distinct groups, and the group-level 
		decisions are 
		combined to make a final decision. For 
		instance, 
		in presidential elections in the United 
		States, 
		the electoral votes of each state is 
		divided 
		between the candidates based on the 
		statewide 
		popular vote, and the candidate with the 
		most 
		electoral votes is chosen for president. 
		As 
		another example, in elections for 
		national 
		parliaments, states or prefectures elect 
		representatives, who then collectively 
		make policy 
		decisions through legislative voting.

		Participation decision in two-tier voting is more complex than in direct voting, since a voter must consider both his influence on his group's decision and the group's influence on the social decision. Given the complexity of the problem, actual voter behavior may differ significantly from theoretical predictions. Moreover, turnout may depend on the electoral rules that specify how votes are aggregated within groups. The rules then affect voter welfare not only directly by converting a configuration of votes into a social decision, but also indirectly by affecting the incentives of voters to participate. To understand how two-tier voting systems work, we extend the standard costly voting model to a multi-group setup, compare voter turnout in theory and experiment, and draw welfare implications for alternative electoral rules.

		We construct a model of two-candidate election with three groups of voters (e.g., states), each having a voting weight (e.g., electoral votes) proportional to population. Each voter decides whether to vote for her preferred candidate
		or abstain, by comparing the expected benefits and costs of voting.
		Each group's weight
		is then allocated to the candidates
		according to some aggregation rule
		based on their vote shares
		within the group.
		A candidate wins if he
		obtains a majority of the total weights across the three groups.
		We consider two alternative
		rules used widely in real politics.
		Under the \textit{winner-take-all rule (WTA)},
		the whole weight of 
		a group
		goes to the candidate who
		receives a majority of votes in the group.
		Under the \textit{proportional rule (PR)},
		each group's weight
		is divided between the candidates
		proportionally to their vote shares
		in the group.

		In our experiment, we simplify the decision problems to reduce
		the complexity of considerations
		necessary for subjects to play the voting game with a multi-group structure.
		Precisely,
		among the three voter groups,
		one group
		(``human group'') consists
		of human subjects
		while the other two groups
		(``computer groups'')
		consist of automated voters programmed to
		play the equilibrium
		strategies.

		We find two prominent behavioral patterns, the {behavioral bandwagon effect} and the {Titanic effect},\footnote{
The former refers to the incentive for the crowd willing to jump on the bandwagon playing lively music, while the latter corresponds to the incentive for the crowd willing to jump off a sinking ship \citep{irwin2000bandwagons}.
        } and that the magnitudes of these effects differ under WTA and PR. The \textit{behavioral bandwagon effect} refers to the case where the observed turnout rate is higher than the theoretical prediction among voters in the majority camp (i.e., those voters in the human group who support the candidate preferred by the majority of voters in the group).\footnote{
We use the adjective \textit{behavioral} to emphasize our focus on the deviation of the observed turnout  from 
the theoretical prediction.}
On the other hand, the \textit{Titanic effect} refers to the case where the observed turnout rate is lower than the theoretical prediction among voters in the minority camp.
		The behavioral bandwagon and Titanic effects are observed under both WTA and PR, but the magnitude is more substantial under PR. 
		Moreover, these effects appear discontinuously around the fifty-fifty split in the support rate between the two candidates, suggesting that \textit{being in the majority or minority itself} affects the participation decision.\footnote{\citet[p. 554]{Barnfield2020} defines the bandwagon effect as ``a phenomenon characterized by a positive individual-level change in vote choice or turnout decision towards a more popular or an increasingly popular candidate or party, motivated initially by this popularity.''}
		Consequently, majority turnout tends to exceed minority turnout, which is consistent with a number of previous experimental studies, yet contrasts with the underdog effect observed by \cite{levine2007paradox}.

		Our theoretical model enables us to simulate the impact of these behavioral effects on voter welfare. 
		The experimental observation that 
  turnout increases among majority voters and decreases among minority voters
  leads to higher (resp. lower) voting costs and larger (smaller) expected benefits from the victory for the majority (minority) candidate, compared with the theoretical prediction.
		According to our model, the impact of expected benefits dominates the effect of voting costs. As a result, the majority’s welfare increases and the minority’s welfare decreases from the theoretical levels.
        We also find that these welfare effects are stronger under PR than under WTA. 
        Therefore, the distribution of voter welfare becomes more unequal than theoretically predicted, and this welfare effect is stronger under PR than under WTA.
        {These observations point to the importance of taking behavioral effects into account in normative evaluation of two-tier election rules.} 
        

		The multi-group framework also allows us to distinguish the majority-minority relationships at the group level and the social level. We call a voter \textit{local majority} if she prefers the candidate supported by a majority of her group, and \textit{global majority} if she prefers the candidate supported by a majority of the whole electorate. Our analysis of the experimental data reveals that being local majority or minority has stronger effects on voter turnout than being global majority or minority does.
		
		\subsection{Related Literature} \label{subsec:lit}

        To our knowledge, this paper is the first to experimentally study endogenous turnout in two-tier voting. Welfare properties of two-tier voting rules have been studied extensively 
         \citep[e.g.][]{BarberaJackson2006, Koriyamaetal2013,Kurz2017,kikuchi2023winner}. However, little attention has been paid as to how those rules affect voters' turnout incentives.\footnote{An exception is \cite{KoriyamaWang2024} which considers a model of two-tier voting with endogenous turnout to examine minority protection under the winner-take-all rule and proportional rule.}
         In particular, \cite{kikuchi2023winner} compares welfare between the winner-take-all rule and the proportional rule, assuming that all voters vote. The present paper complements that study by extending the model to allow voters to abstain.

There are experimental studies that compare turnout under different electoral rules, focusing on the case of direct voting. 
These studies compare different \textit{power sharing rules} (mostly \textit{majority rule} and \textit{proportional representation}\footnote{In the literature, majority rule and proportional representation are defined for direct voting of a single electorate. These power sharing rules should be distinguished from what we call the winner-take-all rule (WTA) and the proportional rule (PR), which are defined for two-tier voting in a multi-group setting.}), assuming that the resulting power shares of parties enter directly into voters' payoffs.
\cite{SchramSonnemans1996} compares voter turnout between majority rule and proportional representation to observe higher turnout under the majority rule, as predicted by the theory. 
\cite{herrera2014turnout} compares turnout between the two rules according to the minority's size. Their data supports the theoretical predictions. \cite{kartal2015laboratory} compares minority representation as well as voter turnout between the two rules. She observes that proportional representation does not improve the representation of a small minority compared with the theoretical prediction.
The main departure of the present paper from the approach of these studies is to define electoral rules as part of a two-tier voting system, whose ultimate outcome is a social decision (e.g., the winner of a presidential election or the result of a legislative vote), not the power shares of parties per se.

\cite{kartal2015laboratory} reports an effect similar to the Titanic effect observed in our experiment. In her experiment, the turnout rate of minority voters under proportional representation was significantly lower than theoretically predicted. As a possible explanation, she suggests that minority voters may have been discouraged from voting, due to an election threshold resulting from a discrete nature of proportional representation. The proportional rule in our experiment has no such election threshold and yet exhibits a similar effect on minority turnout.

The behavioral bandwagon and Titanic effects are closely related to the \textit{bandwagon effect}, which refers to the phenomenon that information that a candidate is more (less) likely to win stimulates (discourages) participation from the supporters of the candidate.\footnote{\cite{MortonOu2015} call
such effects \textit{bandwagon abstention effects}, and distinguish them from \textit{bandwagon vote choices} which refer to vote switches from the likely loser to the likely winner.} The bandwagon effect is not consistent with the underdog effect predicted by standard voting models such as \citeauthor{PalfreyRosenthal1983} (\citeyear{PalfreyRosenthal1983}, \citeyear{palfrey1985voter}) and observed experimentally by \cite{levine2007paradox}. But it
has been observed in many of the previous experiments, including \cite{faravelli2020costly} who reexamined \citeauthor{levine2007paradox}'s (\citeyear{levine2007paradox}) results through Amazon’s Mechanical Turk with larger electorate size and real-effort costs of voting. 

\cite{grosser2010public} experimentally shows that majority voters turn out more often than minority voters, and the release of opinion polls further increases this difference. 
\cite{agranov2018makes} develops a novel experimental design which elicits subjects' beliefs about the election outcome, and shows that the subjects are more likely to vote when they believe that their preferred candidate is more likely to win.
\cite{MortonOu2015} surveys the psychological and political-economy literature on the mental process behind the bandwagon effect. They also examine experimentally the condition under which other-regarding and non-other-regarding bandwagon behaviors are likely to appear, respectively. 
\cite{grillo2017risk} provides a theoretical explanation for the bandwagon effect in terms of risk aversion of voters.


		
		\section{Model} \label{sec:theory}

		\subsection{Two-tier election}
		
		Two candidates, $I=A,B$,
		compete in
		an election.
		The electorate consists of
		three groups, 
		$g=1,2,3$.
		Let $n_g$ denote the population of group $g$.
		Each group $g$ has a
		voting weight (e.g., electoral votes) equal to its population $n_g$.
		The election proceeds as follows.
		Each voter casts one vote
		for a candidate or abstains.
		The weight of each group is then allocated
		to the candidates 
		based on their vote shares in the group, according to some aggregation rule.
		A candidate
		wins the election
		if she receives a majority of the total weights across the three groups.

		Each voter in group $g$
		independently prefers candidate $A$
		with probability $p_g\in[0,1]$
		and candidate $B$ with probability
		$1-p_g$. The probability $p_g$ represents the support rate for candidate $A$ in group $g$.
		The support rate may vary
		across the groups,
		reflecting group-specific bias.
		Each voter obtains a benefit of $\beta>0$
		if her preferred candidate wins.
		Voting incurs a cost $c_i$
		for voter $i$.
		We assume that the voting cost
		is independent across voters
		and uniformly distributed on an interval 
		$[0,\bar{c}]$, which is common knowledge among voters.
		The realized preferences
		and costs
		are private information:
		voter $i$
		knows her preferred candidate and cost $c_i$,
		but not those of the other voters.

		This situation constitutes
		a voting game in which
		voters in all groups
		simultaneously choose
		whether to vote for their preferred candidate
		or abstain, after observing
		their own voting cost and
		preferred candidate.\footnote{Since only two candidates run, voting for her less preferred candidate is a dominated strategy for every voter. Hence, we omit such a choice from our theoretical analysis and experimental setting.}
		Voter $i$ chooses her
		action to maximize
		her expected payoff,
		defined as the benefit
		$\beta$
		weighted by the probability that her preferred candidate wins, minus the cost
		($c_i$ or 0 depending on
		whether she votes or abstains).

		We consider two alternative rules that
		specify how the weight of a group
		is allocated to the candidates
		based on the groupwise voting result.
		Under the \textit{winner-take-all rule (WTA)},
		each group allocates the whole weight
		to the candidate
		who receives a majority of votes in the group.
		Under the \textit{proportional rule (PR)},
		each group divides the weight
		between the two candidates
		proportionally to their vote
		shares in the group.
		For simplicity, we focus on the case where all groups employ the same rule.

		We call a specification of the group sizes 
		$(n_1,n_2,n_3)$ and the candidate support 
		rates $
		(p_1,p_2,p_3)$ a \textit{voting 
		configuration}. We 
		also call a pair of a voting configuration 
		and a 
		rule (WTA or PR) a \textit{voting 
		situation}.

		\subsection{Equilibrium}
		\label{subsec:eqm}
		
		Our theoretical prediction
		for this voting game
		is based on the concept of
		\textit{quasi-symmetric equilibrium}.\footnote{The definition
			below
			is based on the definition
			of (quasi-)symmetric equilibrium
			by \cite{palfrey1985voter}
			and \cite{levine2007paradox},
			which assumes that voters
			preferring the same candidate play the same strategy;
			our definition adapts the original
			concept to the current
			multi-group setting by allowing voters in different groups
			to play different strategies.}  
		In equilibrium,
		each voter maximizes her expected payoff
		given the other voters' strategies.
		By quasi-symmetric,
		we mean
		that all voters in the same
		group play the same strategy.
		It is easy to check that
		any quasi-symmetric equilibrium
		consists 
		of cutpoint strategies.
		A \textit{cutpoint strategy}
		for voter $i$ is a pair of
		cutpoints $\hat{c}_i=(\hat{c}_{i,A},\hat{c}_{i,B})\in
		[0,\bar{c}]^2$
		such that,
		conditional on preferred candidate $I$,
		she votes for $I$ if $c_i\le\hat{c}_{i,I}$
		and abstains if $c_i>\hat{c}_{i,I}$.
		In a quasi-symmetric equilibrium,
		voters in the same group
		have the same cutpoints; hence the equilibrium
		is a profile 
		$\hat{c}=(\hat{c}_{g})_{g=1,2,3}$
		in which $\hat{c}_{g}=(\hat{c}_{g,A},\hat{c}_{g,B})$
		represents
		(with a slight abuse
		of notation) the strategy for
		voters in group $g$.
		Given the cutpoint strategy
		profile $\hat{c}$,
		the probability that a supporter
		of candidate $I$ in group $g$ votes
		equals ${t}_{g,I}=\frac{\hat{c}_{g,I}}{\bar{c}}$,
		which approximately
		equals the {turnout rate}
		among $I$-supporters in group $g$.

		An equilibrium condition can be
		obtained as a set of equations
		involving the probability
		that a voter is \textit{pivotal},
		meaning that her vote
		overturns the election outcome.
		Consider a voter $i$ in group $g$
		who prefers candidate $I$,
		and
		let $\pi_{g,I}(\hat{c})$ be the 
		probability that the voter is pivotal,
		which is a function
		of the cutpoint strategy profile $\hat{c}=(\hat{c}_{g})_{g=1,2,3}$.
		The voter's expected benefit
		of voting is $\beta \pi_{g,I}(\hat{c})$,
		so she votes if $c_i\le \beta \pi_{g,I}(\hat{c})$
		and abstains if $c_i>\beta \pi_{g,I}(\hat{c})$.
		Therefore,
		voter $i$'s choice of the
		cutpoint $\hat{c}_{g,I}$
		is a best response to the profile $\hat{c}$ if and only
		if
		\begin{equation}
			\hat{c}_{g,I}=\beta \pi_{g,I}(\hat{c}).
			\label{eq:br}
		\end{equation}
		The strategy profile $\hat{c}$
		is an equilibrium
		if and only if equation (\ref{eq:br})
		holds for all groups $g=1,2,3$ and 
		candidates $I=A,B$.
		The two rules, WTA
		and PR,
		induce different
		pivot probability functions $\pi_{g,I}(\cdot)$
		and, therefore, different
		equilibria.

		\section{Experimental Design}

		\subsection{Human and computer groups}	
		
		We conducted an experiment on the voting 
		game described in the previous section. To 
		reduce 
		the complexity of the decision problem for 
		the 
		subjects, we set up only group 1 as a 
		\textit{human group} consisting of 
		subjects, while the remaining
		groups 2 and 3 as \textit{computer groups} 
		consisting of automated voters programmed 
		to play 
		the equilibrium strategies.

		\subsection{Decision problem}

\begin{figure}[t]
	\centering
	\includegraphics[width=\textwidth]{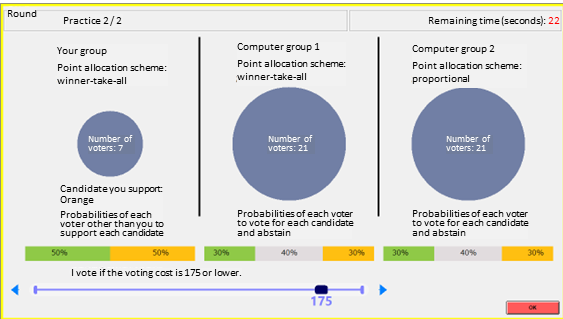}
	\caption{The decision screen}
	\label{ztuenv}
\end{figure}

Figure \ref{ztuenv} is an image of the screen shown to the subjects in each voting situation.
Two candidates are labeled Orange and Green.\footnote{Words with political connotations, such as party, left or right, Electoral College, and colors red or blue, are intentionally avoided so that subjects' behavior is not affected by a particular political orientation.} 
Voters in the three groups independently become a 
supporter of either candidate, according to the 
probabilities $(p_1, p_2, p_3)$. The 
preferences are private information: each subject 
knows her own preferred candidate, but does not 
know the preferences of the other voters (human or 
computer).
Each subject observes the population sizes ($n_1, n_2, n_3$), the aggregation rule (WTA or PR), the support rates of the candidates ($p_1$) in the human group, and the probabilities of each automated voter in the computer groups to vote for Orange, vote for Green, or abstain, where these probabilities for the computer groups are derived from the equilibrium strategies.

Given this information, each subject chooses a threshold of the cost below which she votes, by moving the slider over the interval displayed on the screen.\footnote{		
\cite{levine2007paradox} designed their voting experiment based on the incomplete-information game with privately known costs determined randomly for each voter (\citeauthor{palfrey1985voter}, \citeyear{palfrey1985voter}). Their data supported the comparative statics of the theory for WTA. The advantage of the incomplete-information game is the uniqueness of equilibrium; each voter employs the cutoff strategy under which she votes if her voting cost is lower than or equal to a threshold and abstains otherwise. \cite{grosser2020voting} surveys this type of voting experiment with incomplete information.
\cite{aguiar2016experimental} designed their referendum experiment using this strategy-method approach.} The subject can choose any integer in the set $S=\{ 0, 1, 2, \dots, 200\}$.

Once all subjects have made their choice,
the voting costs are independently drawn from the uniform distribution over $S$. 
Each voter votes for her preferred candidate if the cost is within her threshold, and abstains otherwise.
The votes are aggregated within each group, and the voting weights are allocated to the candidates according to the pre-specified rule, which determines the winner of the election.

Each subject obtains 1,000 points if her preferred candidate wins. The voting cost is subtracted if she has voted. The screen then displays a summary of the results: the number of votes cast for each candidate, the amount of each group's voting weight allocated to each candidate, the election winner, the points obtained by the subject, and the subtracted cost.

		\subsection{Local and global majority}
{Table \ref{table:config} shows the 18 voting configurations used in our experiment.}
Given our setting with three groups of voters, it is 
useful to define the notions of majority and minority in local and global senses.
We say that candidate $A$ is \textit{local 
minority} (and 
candidate $B$ \textit{local majority}) if $A$ is 
minority in the human group, i.e., if 
$p_1<0.5$.
We say that candidate $A$ is \textit{global 
minority} (and 
candidate $B$ \textit{global majority}) if $A$ is 
minority among the entire electorate, i.e., if 
$\bar{p}<0.5$, 
where $\bar{p}$ denotes the overall support rate
for $A$ among the electorate:
$\bar{p}=(p_1n_1+p_2n_2+p_3n_3)/(n_1+n_2+n_3)$.

\input{Table_voterconfig}

Our experiment focuses on those voting 
configurations in which one candidate, say 
candidate $A$, is weak minority both locally and 
globally (i.e., $p_1\leq0.5$ and $\bar{p}
\leq0.5$).
The 18 voting 
configurations used in the experiment are 
classified
into four \textit{categories}, as in Table 
\ref{table:category}. In category 1, called 
impartial culture (\textit{IC}), the two candidates are 
equally 
preferred by voters locally and globally, and 
hence there is no majority or minority candidate 
(i.e., $p_1=\bar{p}=0.5$). 
Category 2 concerns global asymmetry only (hereafter \textit{Global}): 
the two 
candidates are equally preferred in group 1, but 
candidate $B$ is more preferred among the whole 
electorate (i.e., $p_1=0.5$ and $\bar{p}<0.5$), 
implying that candidate $B$ is preferred over 
candidate $A$ in groups 2 and/or 3. 
Category 3 concerns the reverse case, local asymmetry only (hereafter \textit{Local}). Candidate 
$B$ is more preferred in group 1, but the two 
candidates are equally preferred among the whole 
electorate (i.e., $p_1<0.5$ and $\bar{p}=0.5$), 
implying that candidate $A$ is preferred over 
candidate $B$ in groups 2 and/or 3. 
Finally, category 4 concerns both local and global asymmetry (hereafter \textit{Both}): 
candidate $B$ is preferred over 
candidate $A$ locally and globally (i.e., 
$p_1<0.5$ and $\bar{p}<0.5$).

\begin{table}[h]
	\begin{center}
		\caption{Categories of
			voting configurations.}
		\label{table:category}
			\begin{tabular}{lcc}
				
				\hline
				Category & Local support rate & 
				Global support rate\\
				\hline
				1. Impartial Culture (IC) & 
				$p_1=0.5$ & $\bar{p}=0.5$\\
				2. Global asymmetry only (Global) & 
				$p_1=0.5$ & $\bar{p}<0.5$\\
				3. Local asymmetry only (Local) & 
				$p_1<0.5$ & $\bar{p}=0.5$\\
				4. Both local and global asymmetry (Both) 
				& $p_1<0.5$ & $\bar{p}<0.5$\\
				\hline
			\end{tabular}
	\end{center}
	\label{tab:category}
\end{table}

		\subsection{Sessions}
		We conducted the experiment through 8 sessions, 
		each with 21 human subjects. 
  Each session consists of 36 rounds, with all subjects playing each of the 18 voting configurations exactly once under each of WTA and PR.
  The voting situations are divided into 
		four {blocks} specified by the electoral 
		rule (WTA 
		or PR) and the size of the human group (7 
		or 21). 
		We changed the order of blocks
  {and reversed the order of voting configurations within a block session by 
		session 
		to minimize the order effect (see Table 
		\ref{tab:sessions}).}

		\begin{table}[ht] 
		\centering
		\caption{Session summary: order of blocks}
		\begin{tabular}{@{}lllll@{}}
			\toprule
			Session No. & 1,5         & 2,6         & 3,7         & 4,8          \\ \midrule
			1st block       & WTA, 21 & WTA, 7  & PR, 21  & PR, 7     \\
			2nd              & WTA, 7  & WTA, 21 & PR,7    & PR,21    \\
			3rd              & PR, 21  & PR, 7   & WTA, 21 & WTA, 7     \\
			4th              & PR, 7    & PR, 21   & WTA, 7  & WTA, 21  \\ \bottomrule
		\end{tabular} \label{tab:sessions}
	\end{table}

		{
		One typical session lasted for 
		approximately 130 minutes. At the end of 
		each 
		session, one voting situation per block was randomly 
		selected, and subjects were paid for the sum of points earned in four randomly selected situations.    
  We sent Amazon gift card via email for payment.
		The average payment was 3,053 JPY (1 point = 
		1 JPY, 27.9 USD as of June 1st, 2021).}

		\subsection{Logistics}

	The eight sessions took place in March and June 2021. Our experiment was programmed with the z-Tree Unleashed \citep{duch2020z} and ran online through a virtual server in Amazon Elastic Compute Cloud (EC2). Subjects were recruited via the ORSEE \citep{greiner2015subject} from the campus-wide student subject pool at Osaka University. Thirty-four percent of the participants were female.

		Prior to each session, subjects were gathered in a Zoom meeting, where they were identified and anonymized. 
		Instructions were 
		shown by sharing a screen, and read by a 
		text-to-speech software, which served to control time and tone. 
		While the instructions were being read, an experimenter 
		indicated the relevant part of the screen by a pointer. 
		The 
		instruction material was accessible for 
		the 
		subjects anytime during the session  via a 
		hyperlink.\footnote{%
			The full experimental instructions, including screenshots and the questionnaire, are provided in Online Appendix. 
		} 
            {After playing 36 rounds of voting, the subjects answered the post-experimental questionnaire distributed by Google Forms (See Section 4.3 for the details).}

		\section{Results}

		\subsection{The bandwagon and Titanic effects}

{Table \ref{tab:comparison_turnout} shows the comparison of turnout rates between equilibrium and experiment. The turnout rate ${t}_{1,I}^{ R}$ represents that of a voter in the human group (i.e., group 1) for each candidate $I\in\{A,B\}$ and rule $R\in\{\rm WTA, PR\}$, over the 18 voting configurations used in our experiment. }

\input{table_turnoutrates}

\begin{figure}[t]
			\centering
			\includegraphics[width=0.495\textwidth]{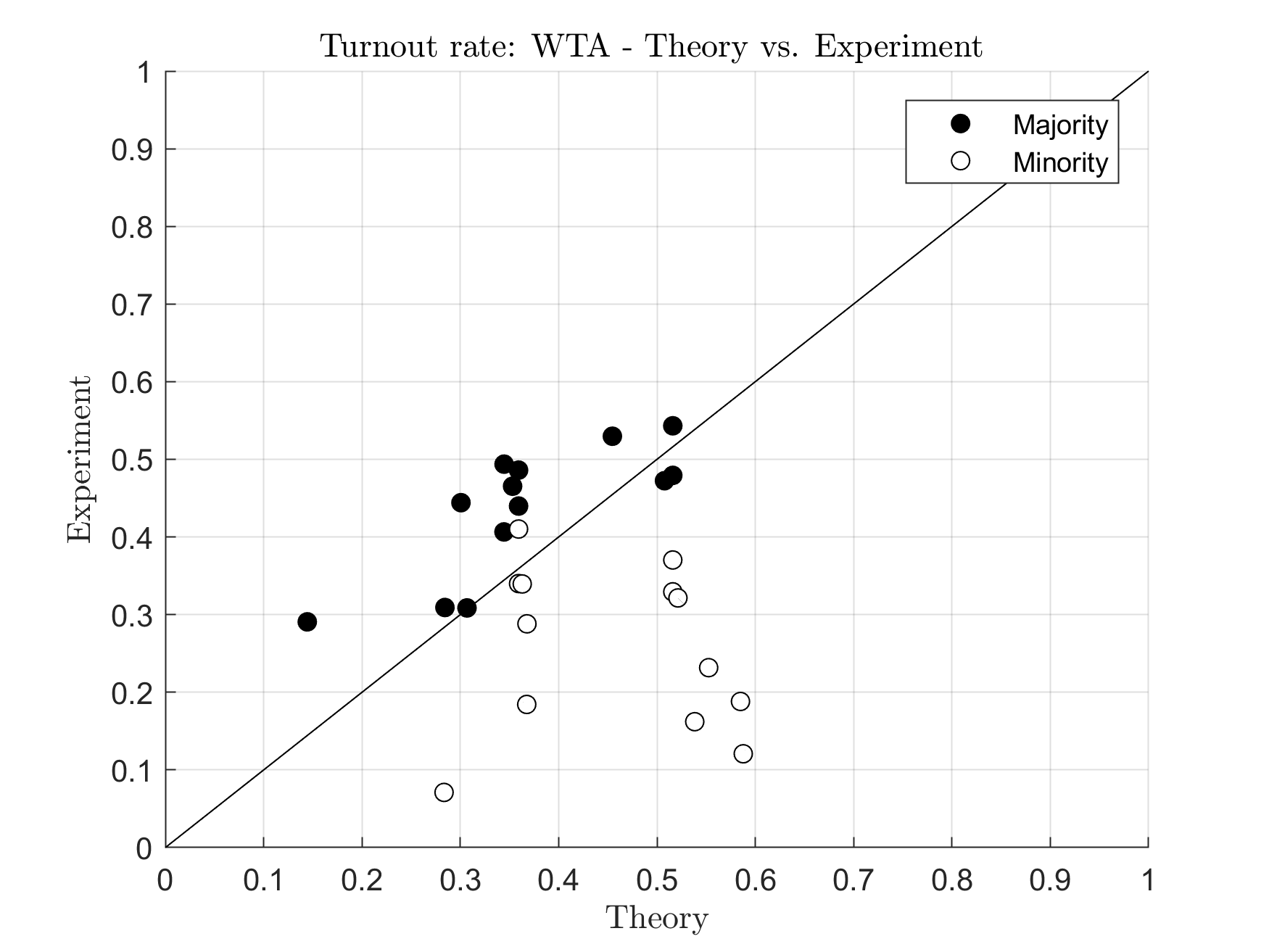}
			\includegraphics[width=0.495\textwidth]{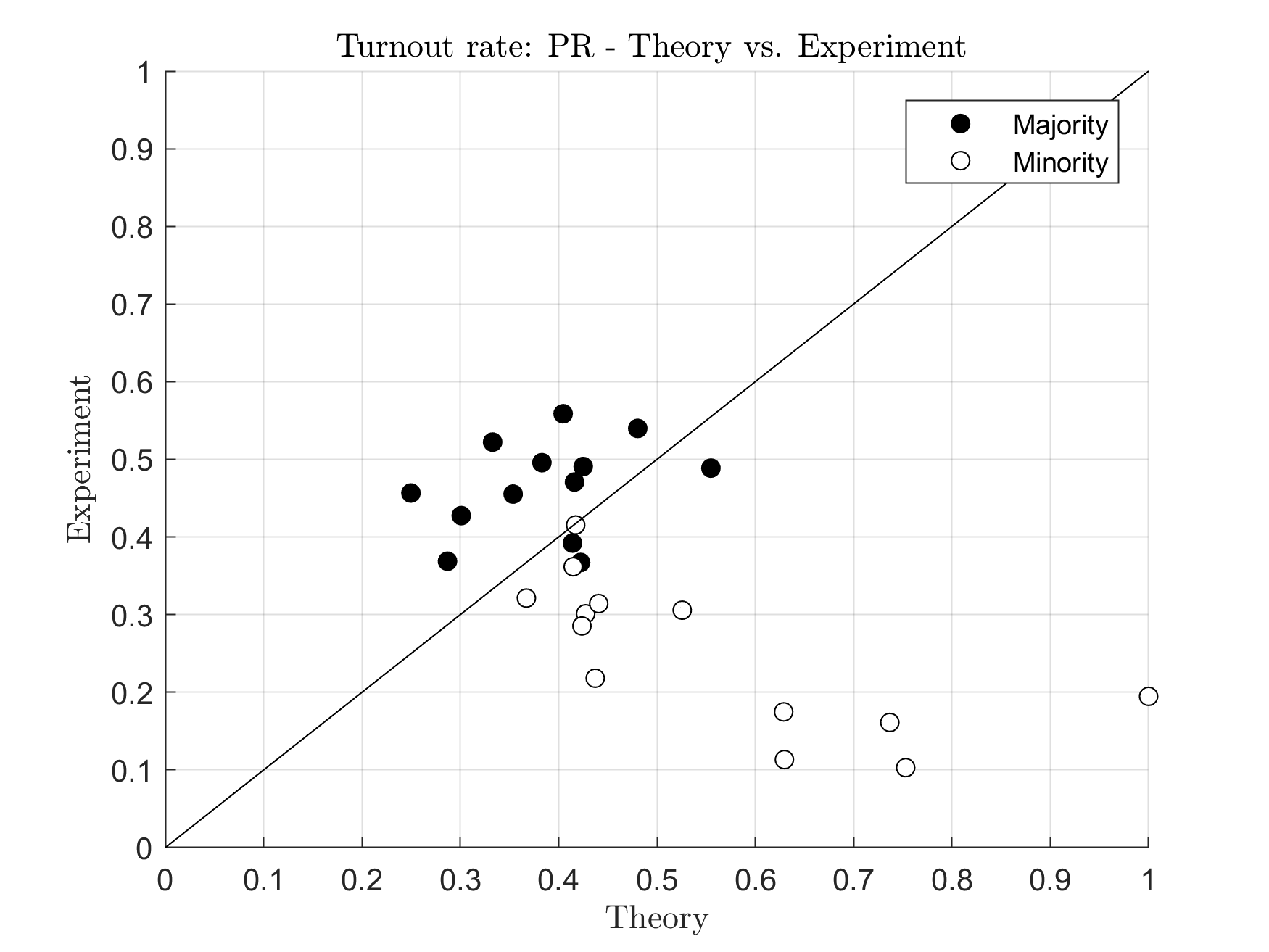}
			\caption{Turnout rate: comparison between theory and experiment.}
			\label{fig:PrVote_TheoExpe}
		\end{figure}

			We first investigate the effects of being majority or minority on subjects' participation decisions. We here focus on the voting configurations in the categories where the majority and the minority are well-defined (Global, Local and Both). 
   Figure \ref{fig:PrVote_TheoExpe} plots the pairs of theoretical and observed turnout rates in the majority camp ($\bullet$) and the minority camp ($\circ$) in the human group under WTA (left) and PR (right).\footnote{
   In our experiment, there is no voting configuration in which the local majority and the global majority contradict each other (Table \ref{table:category}), although such a contradiction is mathematically possible. We therefore use the simple expression majority (minority) to designate either local, global majority (minority) or both.
}  
   Each dot represents a voting configuration displaying the turnout rates averaged over all sessions.

			We find two prominent behavioral patterns, the behavioral bandwagon effect and the Titanic effect, and that these effects are stronger under PR than WTA. We say that the \textit{behavioral bandwagon effect} occurs if the observed turnout rate among voters in the majority camp is higher than the theoretical prediction. On the other hand, we say that the \textit{Titanic effect} occurs if the observed turnout rate among voters in the minority camp is lower than the theoretical prediction. 
   Figure \ref{fig:PrVote_TheoExpe} shows that these effects are evident: most black dots lie above the 45-degree line, while most white dots lie below it.

		The non-parametric one-sample Wilcoxon tests statistically confirm the two behavioral effects. Table \ref{tab:tests_TheoExpe} summarizes the test results. 
		For the behavioral bandwagon (resp. Titanic) effect, out of the 13 voting configurations {in Global, Local and Both} categories 
		the difference between theory and experiment is statistically significant in 4 (10) under WTA and 6 (11) under PR. The Titanic effect appears stronger than the behavioral bandwagon effect, and both effects appear more prominent under PR than WTA.

\input{table7new_BandwagonTitanic}	
  
		\subsection{Welfare consequences}
		\label{subsec:welfare}

What are the implications of these behavioral patterns for voter welfare under different electoral rules? 
To answer this question, we provide a comparison of the expected welfare of the majority and minority voters in each voting configuration based on the experimental data.
Observed threshold values are regarded as constituting a sample from a hypothetical mixed strategy played in each voting situation, and the ex-ante expected payoffs are computed assuming that  all subjects play the mixed strategy.\footnote{We need to construct a mixed strategy rather than simply taking the average of realized payoffs in the experiment because
the realized payoffs are correlated
among subjects in the same session
as they have the same election outcome.
Estimated welfare can then
be interpreted as
the expected payoff
when the subjects in group 1 are
a random sample 
from the population.
In an experimental study of mechanism design, \cite{hoffmann2021flip}
	use a method similar to ours to compare realized 
	payoffs in the laboratory
	under different mechanisms: they
	use the empirical distribution
	of strategies chosen by subjects
	to construct behavioral strategies
	for each type of agent, and then
	compute the payoffs and surplus assuming that
	the agents follow those behavioral strategies.}
The behavior of the two computer groups is fixed at the equilibrium strategies, as is done in the experiment.\footnote{For each voting situation,
we have the two
sets of observed cutpoints: those chosen
by subjects supporting
candidate $I=A,B$.
From the set
of $I$-supporters' cutpoints,
we obtain the empirical
distribution
$\sigma_{1,I}$
of cutpoints,
which allows us to define
the mixed cutpoint strategy
$\sigma_1=(\sigma_{1,A},\sigma_{1,B})$
for the subjects in group 1, called
the \textit{estimated mixed strategy}.
We then define
the \textit{estimated welfare}
for $I$-supporters in group 1 
in the given voting situation
as
the expected value
of their payoff
when all voters in group 1
play the mixed strategy $\sigma_1$.}

\begin{figure}[t]
	\centering
	\includegraphics[width=0.495\textwidth]{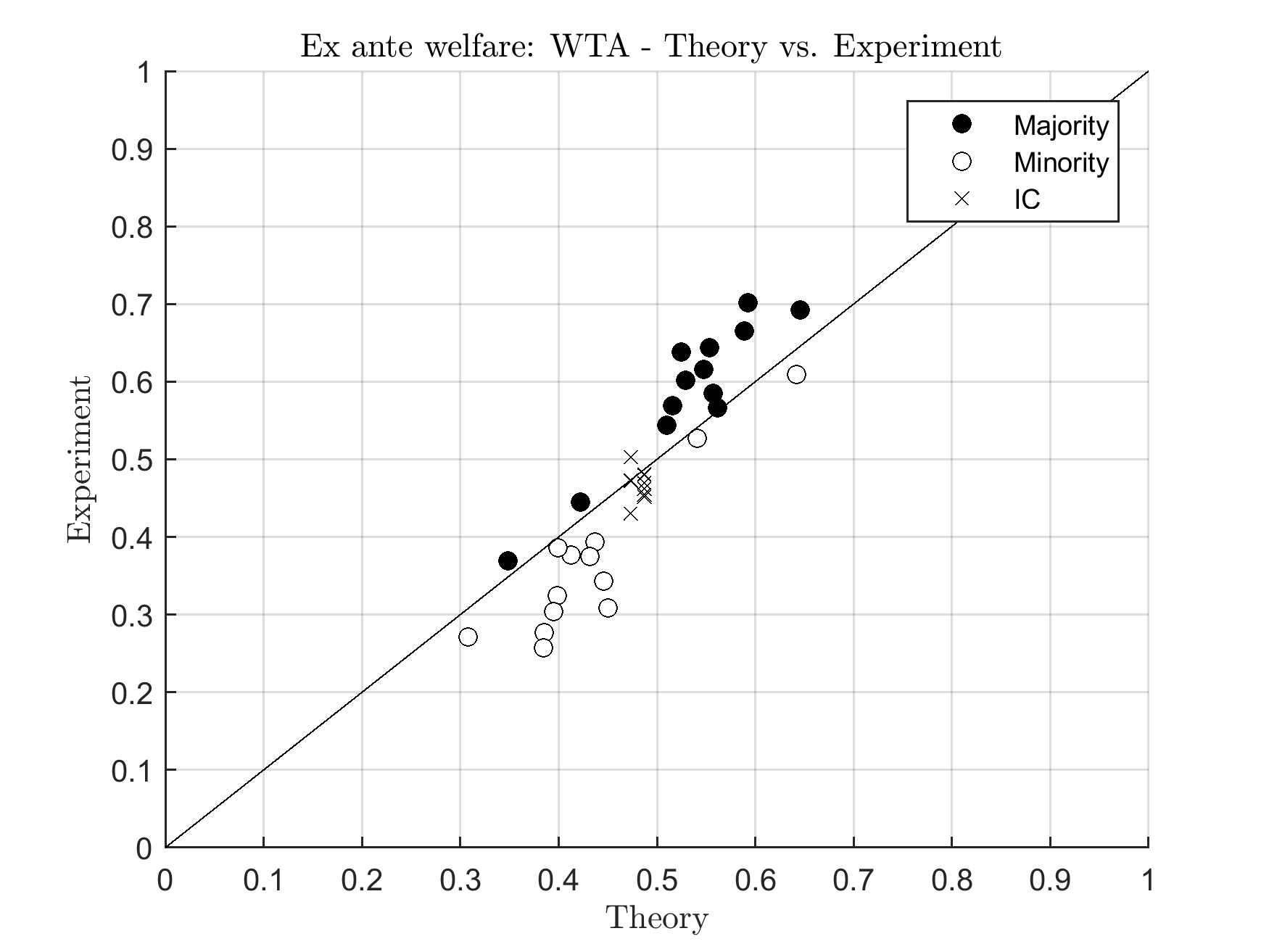}
	\includegraphics[width=0.495\textwidth]{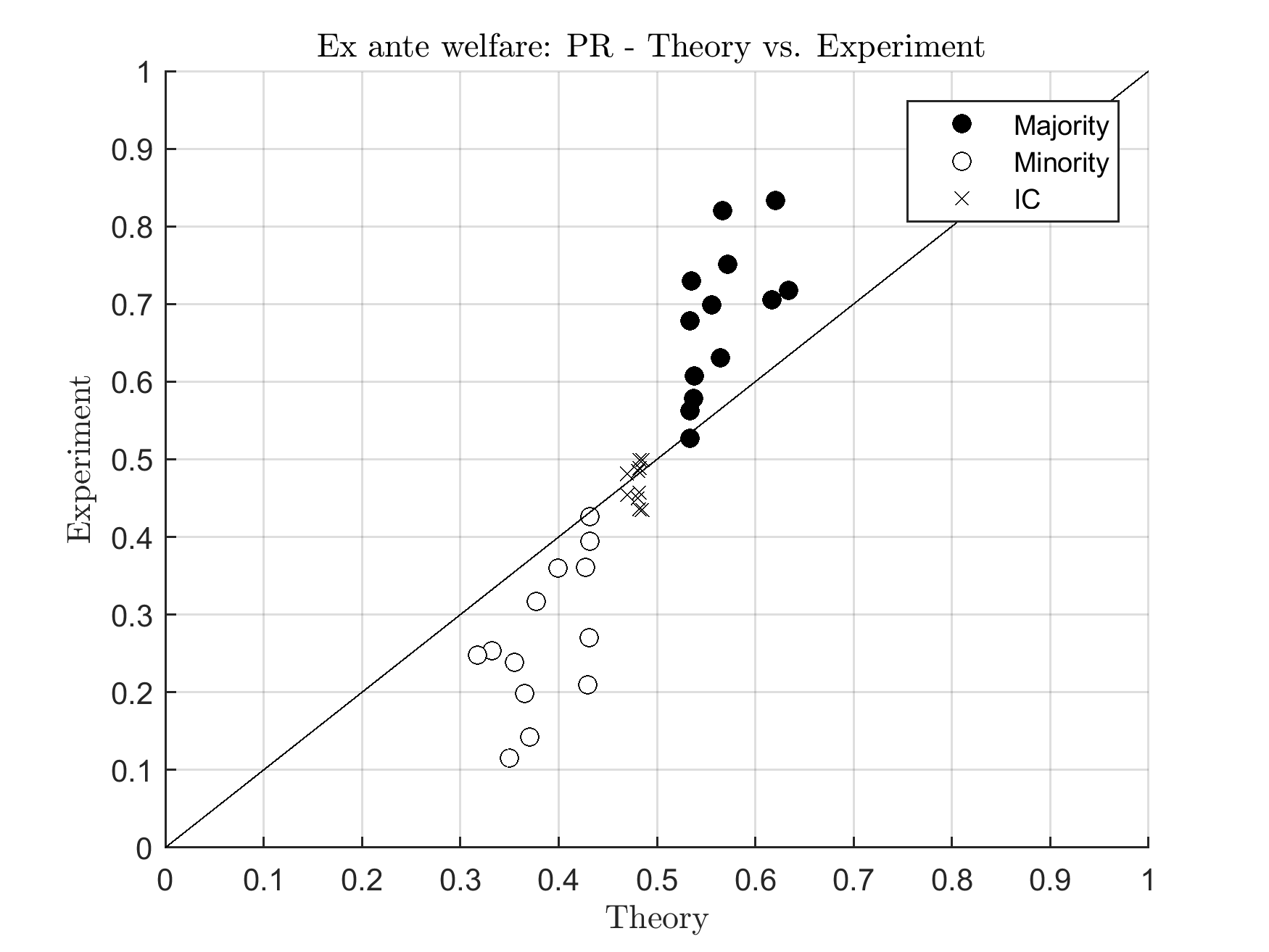}
	\caption{Ex-ante welfare: comparison between the theory and experiment.}
	\label{fig:Wel_TheoExpe}
\end{figure}

\begin{figure}[t]
	\centering
	\includegraphics[width=0.495\textwidth]{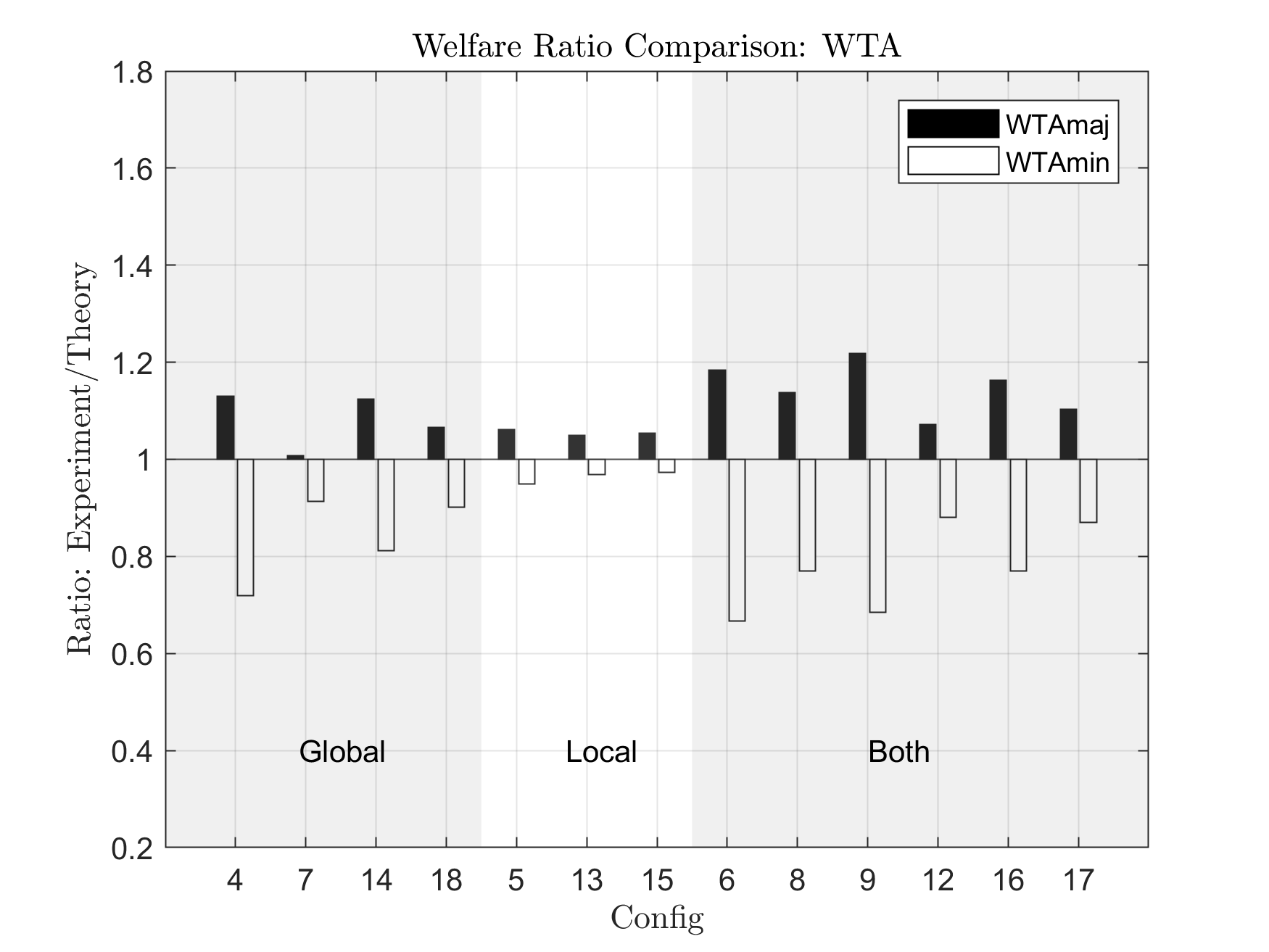}
	\includegraphics[width=0.495\textwidth]{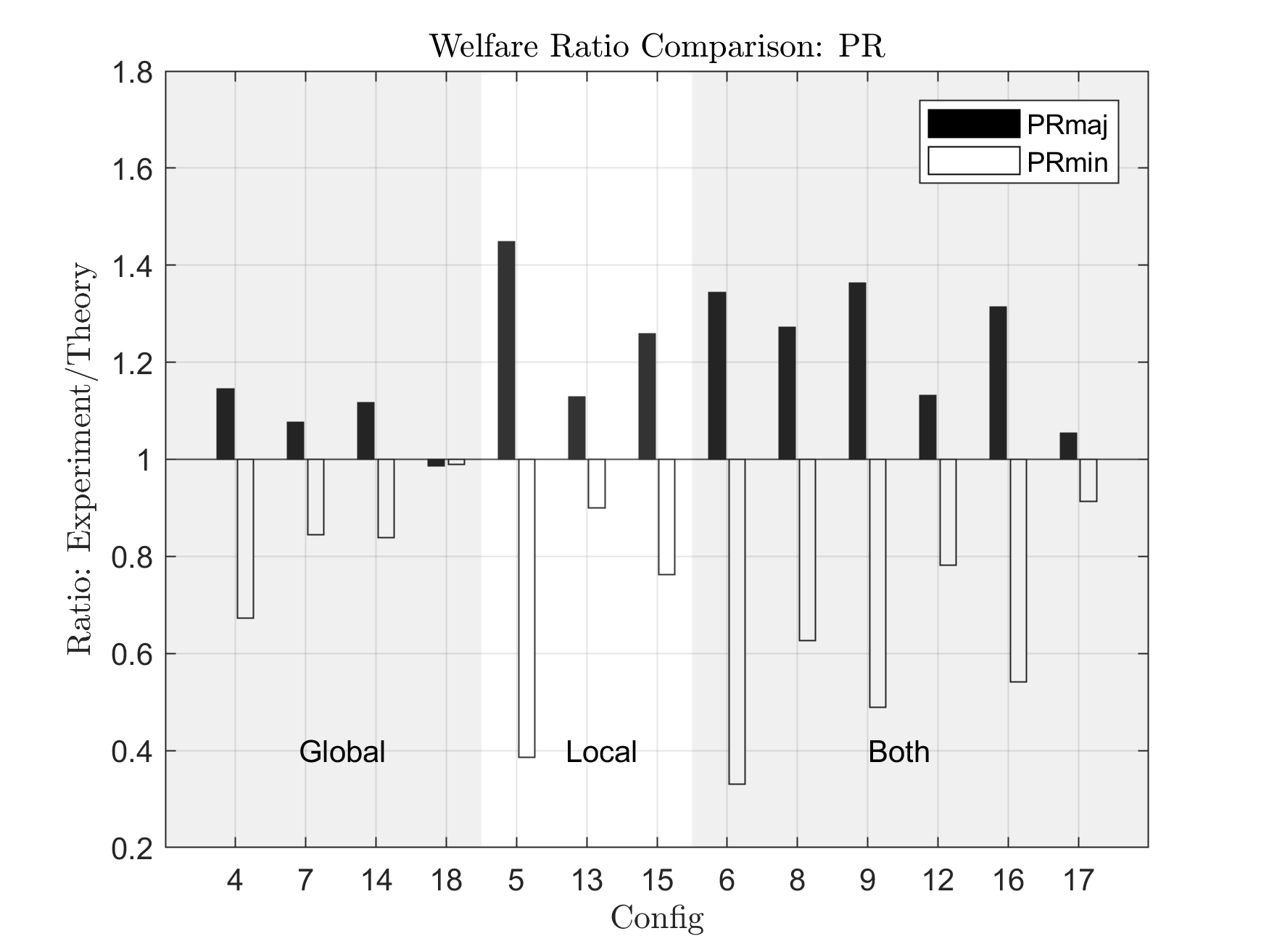}
	\caption{Ratio of welfare in the experiment to the theory.}
	\label{fig:WelRatio}
\end{figure}

Figure \ref{fig:Wel_TheoExpe} shows the scatter plot of the theoretical and experimental welfare levels for each rule. The vertical (resp. horizontal) axis represents the experimental (theoretical) values. 
We see that the dots for the majority ($\bullet$) lie above the 45-degree line, while those for the minority ($\circ$) lie below it.
This means that the behavioral bandwagon and Titanic effects 
caused an increase in the majority's welfare and a decrease in the minority's welfare. 
The trend is similarly observed under both rules, but its extent is more pronounced under PR.
This pattern is clearly visible in Figure \ref{fig:WelRatio}, which shows the ratio of the experimental welfare level divided by the theoretical welfare level for each configuration.
Under both rules, the majority (resp. minority) bars extend above (below) the ratio value of 1. 
For both the majority and minority camps, the bars are longer under PR than WTA.

It is worth noting that the welfare predictions in the IC categories are remarkably close to those in the experiment, confirming the validity of the model in the IC environment and suggesting that the deviation of welfare from equilibrium arises due to behavioral effects related to being in the majority or minority.


\begin{table}[h]
    \input{table_WelfareComparison}

\label{tab:welfarecomp}
\end{table}

Table \ref{tab:welfarecomp} shows the comparison of expected welfare between theory and experiment, averaged over configurations in each category.
These numbers confirm that
(i) welfare increases in the majority and decreases in the minority, and 
(ii) the differences are larger under PR than WTA in most cases. 
Table \ref{tab:welfarecomp} also shows the ex ante Gini coefficients. 
We observe that inequality increases, except for the Local category under WTA.\footnote{
This exception is due to the fact that global symmetry ($\bar{p}=0.5$) is obtained by asymmetry in human groups ($p_1<0.5$) and inverse asymmetry in computer groups ($p_2,p_3>0.5$). See Table \ref{table:config}. The opposite majority in the computer groups is so dominant under WTA that the expected welfare of local majority is \textit{lower} than that of local minority (Table \ref{tab:welfarecomp}). Consequently, behavioral effects in human groups lead to an increase in welfare among the voters with lower welfare, and thus imply a \textit{decrease} in inequality.
}

				\subsection{Dependence on the local and global support rates}
		
		Figures \ref{fig:localglobalMinMaj} shows how the turnout rate (vertical axis) varies as a function of the local and global support rates for the preferred candidate (horizontal axis) in theory (+) and experiment ($\ast$). 
		Here again, the Titanic effect and the behavioral bandwagon effect are visible. On the left-hand side of 0.5, the observed turnout is below the predicted turnout, and the inverse relationship holds on the right-hand side. These effects appear more clearly concerning the local support rate (the upper figures) than the global support rate (the bottom figures).
		
		\newpage
		
		\begin{midpage}
			
		\begin{figure}[h]
			\begin{tabular}{cc}
				\begin{minipage}[t]{0.45\hsize}
					\centering
					\includegraphics[width=7.5cm, angle=0]{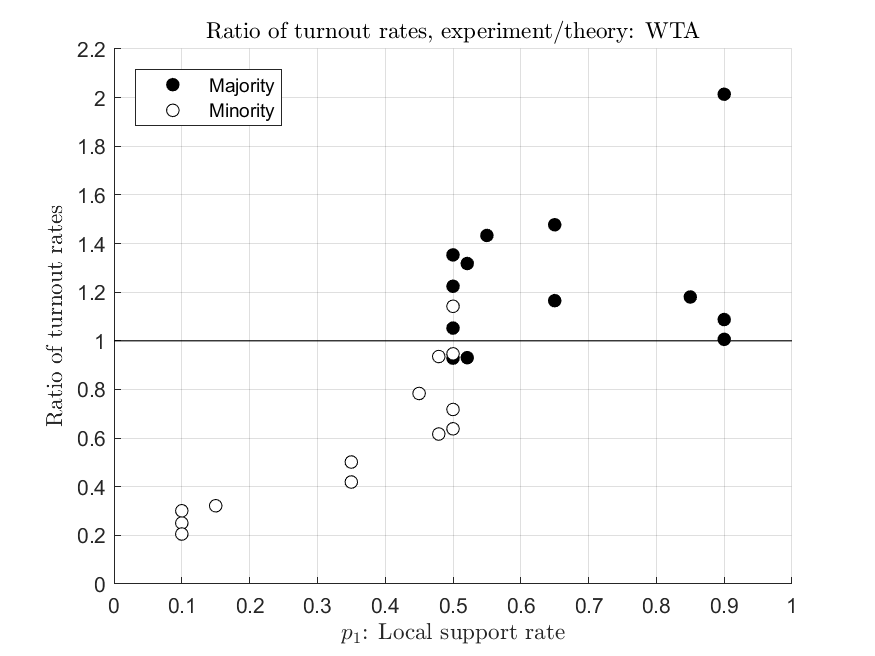}

				\end{minipage} &
				\begin{minipage}[t]{0.45\hsize}
					\centering
					\includegraphics[width=7.5cm, angle=0]{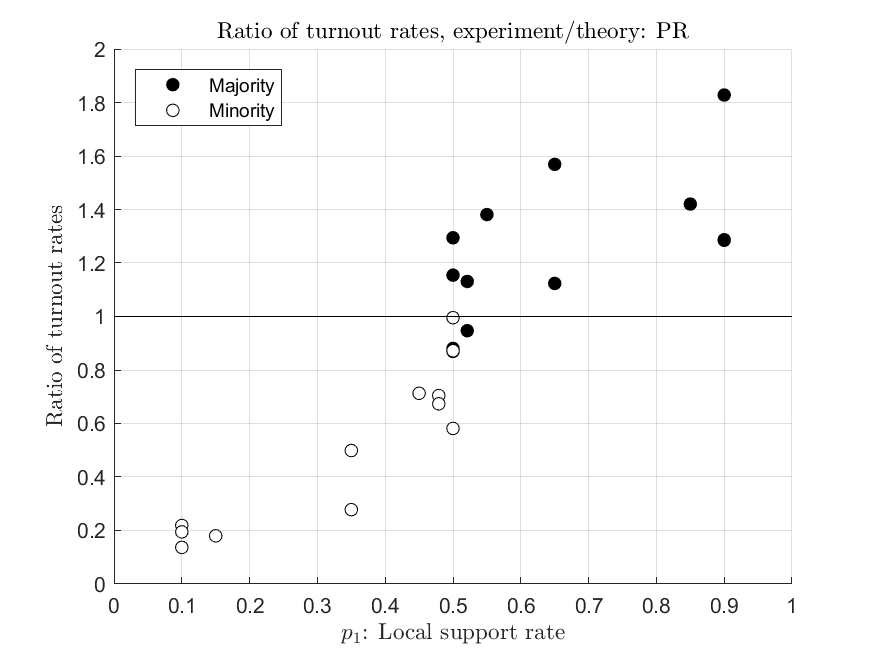}
				\end{minipage} \\
				
				\begin{minipage}[t]{0.45\hsize}
					\centering
					\includegraphics[width=7.5cm, angle=0]{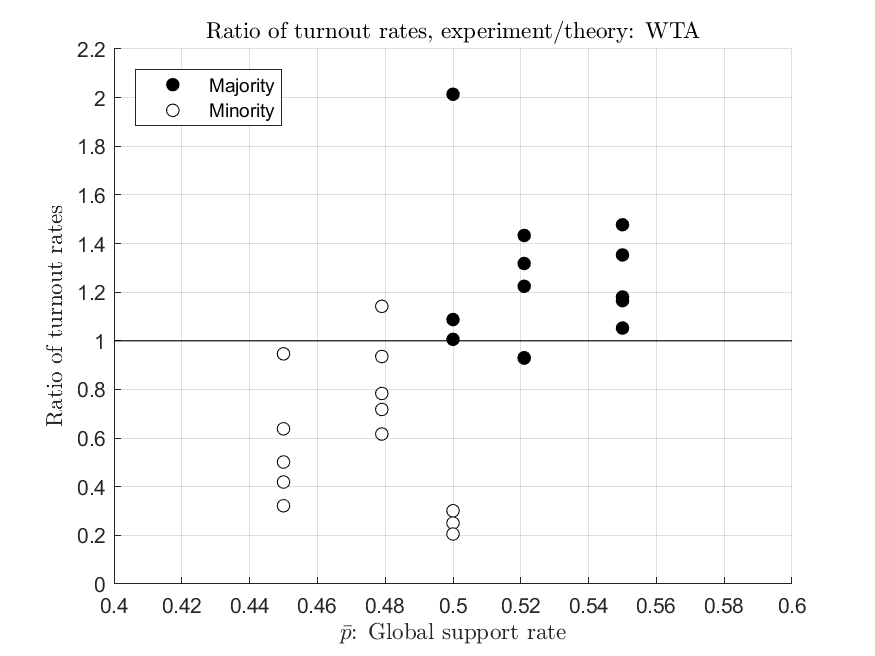}
				\end{minipage} &
				\begin{minipage}[t]{0.45\hsize}
					\centering
					\includegraphics[width=7.5cm, angle=0]{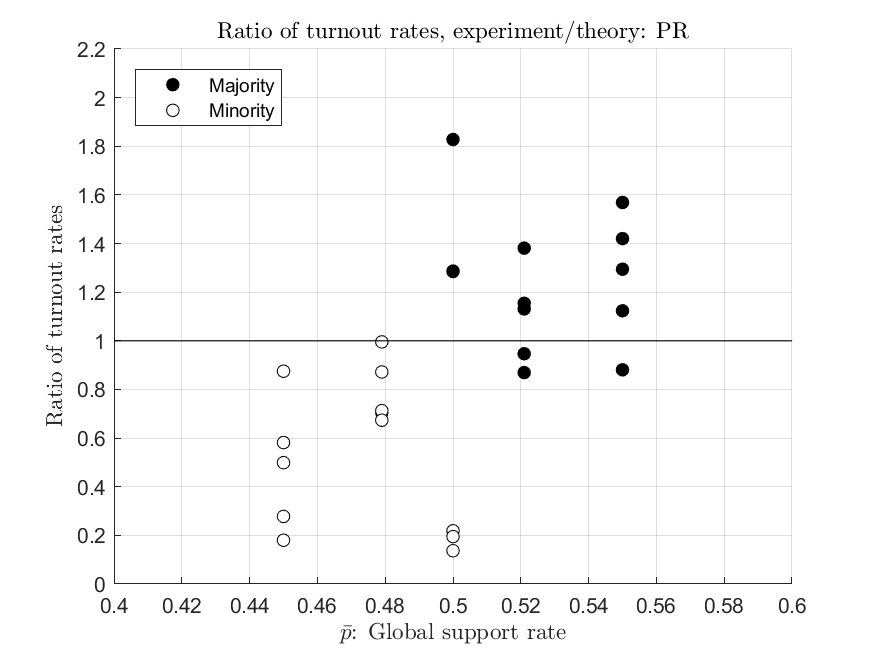}
				\end{minipage} 
			\end{tabular}
			\caption{Ratio of the turnout rates as a function of the local support rate $p_1$ and the global support rate $\bar{p}$.}
			\label{fig:localglobalMinMaj}
		\end{figure}

		\end{midpage}
		
		\newpage

		\subsection{Regression analysis} \label{subsec:tobit}
  We report regression results to further investigate turnout behavior at the individual level (Table \ref{tab:regression_dataonly_and_deviation_controlsomittted}).
Two dependent variables are employed, concerning the voter turnout for each of WTA and PR. 
The first variable, [Experiment], is the normalized threshold chosen by each subject as the maximum cost acceptable for voting, divided by the range of voting costs (i.e., 200). Since the variable is bounded by the interval [0,1], a two-limit Tobit model is applied.
The second dependent variable, [Experiment]$-$[Theory], measures the extent to which each subject’s turnout behavior \textit{deviates} from the equilibrium. A generalized least squares model is employed for this regression.
Random effect models are used for each regression, in order to account for the individual-specific effects of each subject who has made 18 decisions under each rule.

\begin{table}[htbp]\centering
\def\sym#1{\ifmmode^{#1}\else\(^{#1}\)\fi}

\caption{Determinants of turnout, random effects regressions} 
\begin{adjustbox}{max width=\textwidth}
\begin{tabular}{l*{4}{D{.}{.}{-1}}}
\toprule
    &\multicolumn{2}{c}{[Experiment]}&\multicolumn{2}{c}{[Experiment]$-$[Theory]}\\
    &\multicolumn{2}{c}{Tobit}&\multicolumn{2}{c}{GLS}\\
    \cline{2-3}    \cline{4-5}\\
    &\multicolumn{1}{c}{(1) WTA}&\multicolumn{1}{c}{(2) PR}&\multicolumn{1}{c}{(3) WTA}&\multicolumn{1}{c}{(4) PR}\\
\midrule
Local Majority      &      0.0315         &      0.0538\sym{*}  &      0.0081         &      0.0459\sym{**} \\
                    &    [0.0294]         &    [0.0277]         &    [0.0219]         &    [0.0210]         \\
\addlinespace
Local Majority $\times$ Local Majority Share&     -0.5503\sym{***}&     -0.1867\sym{**} &      0.1549\sym{**} &      0.2381\sym{***}\\
                    &    [0.0915]         &    [0.0852]         &    [0.0679]         &    [0.0645]         \\
\addlinespace
Local Minority      &     -0.0721\sym{**} &     -0.1117\sym{***}&     -0.0577\sym{**} &     -0.0555\sym{**} \\
                    &    [0.0333]         &    [0.0315]         &    [0.0242]         &    [0.0235]         \\
\addlinespace
Local Minority $\times$ Local Minority Share&      0.9069\sym{***}&      0.5897\sym{***}&      0.6385\sym{***}&      1.3009\sym{***}\\
                    &    [0.1441]         &    [0.1479]         &    [0.1021]         &    [0.1084]         \\
\addlinespace
Global Majority   &     -0.0477         &     -0.0735\sym{**} &     -0.0178         &     -0.0377         \\
                    &    [0.0389]         &    [0.0364]         &    [0.0290]         &    [0.0276]         \\
\addlinespace
Global Majority $\times$ Global Majority Share&      2.7268\sym{***}&      2.4255\sym{***}&      1.5055\sym{**} &      1.3617\sym{**} \\
                    &    [0.8014]         &    [0.7581]         &    [0.5960]         &    [0.5742]         \\
\addlinespace
Global Minority   &     -0.0413         &     -0.0259         &     -0.0248         &     -0.0229         \\
                    &    [0.0422]         &    [0.0398]         &    [0.0308]         &    [0.0299]         \\
\addlinespace
Global Minority $\times$ Global Minority Share&      1.7620\sym{*}  &      2.7850\sym{***}&      1.7377\sym{**} &      1.3906\sym{**} \\
                    &    [0.9726]         &    [0.9083]         &    [0.7075]         &    [0.6779]         \\
\addlinespace

Local Population          &     -0.0034\sym{***}&     -0.0003         &      0.0088\sym{***}&      0.0114\sym{***}\\
                    &    [0.0013]         &    [0.0012]         &    [0.0009]         &    [0.0009]         \\
\addlinespace
Local Population Share     &      0.2284\sym{***}&      0.3747\sym{***}&      0.1879\sym{***}&     -0.2826\sym{***}\\
                    &    [0.0820]         &    [0.0769]         &    [0.0605]         &    [0.0579]         \\
\addlinespace

Round             &     -0.0092\sym{***}&     -0.0043\sym{***}&     -0.0067\sym{***}&     -0.0030\sym{***}\\
                    &    [0.0011]         &    [0.0010]         &    [0.0008]         &    [0.0008]         \\
\addlinespace
Constant            &      0.5509\sym{***}&      0.3966\sym{***}&     -0.0246         &      0.0044         \\
                    &    [0.1283]         &    [0.1251]         &    [0.0836]         &    [0.0840]         \\
\addlinespace
\midrule
Individual Traits         & \text{Yes} & \text{Yes} & \text{Yes} & \text{Yes} \\
\midrule

Observations        &        3024         &        3024         &        3024         &        3024         \\
Overall model significance &        470.37&        443.86&        818.72&        1286.07\\
\bottomrule
\multicolumn{5}{l}{\footnotesize Standard errors in brackets}\\
\multicolumn{5}{l}{\footnotesize \sym{*} \(p<0.10\), \sym{**} \(p<0.05\), \sym{***} \(p<0.01\)}\\
\multicolumn{5}{l}{\footnotesize
Overall model significance is evaluated by Wald chi square statistic with the degree of freedom of 18.}\\
\end{tabular}
\end{adjustbox}
\label{tab:regression_dataonly_and_deviation_controlsomittted}
\end{table}

Independent variables include those associated with subjects’ decision-making environments and individual characteristics. 
  We also include the round number to capture the time trend of voting behavior. Since we have three groups in our elections (i.e., the human group and two computer groups), each factor of the voting configuration has both within and between-group aspects. 
		
		The majority dummy in the human group (i.e., \textit{Local Majority}) takes the value 1 if each subject belongs to the majority in the human group and 0 otherwise. Note that the majority is defined in the ex-ante sense: it is the camp to which each subject is assigned with a higher probability than the other camp (i.e., $B$ if $p_1 \neq 0.5$). The interaction term between the majority dummy and the share of the majority in the human group (i.e., \textit{Local Majority Share}) expresses how large the majority occupies. The minority dummy (i.e., \textit{Local Minority}) and its interaction term with its share in the human group (i.e., \textit{Local Minority Share}) are defined similarly to those for the majority. Hence, the benchmark is when each subject is assigned to the two camps equally likely (i.e., $p_1=0.5$).\footnote{{The share variables are therefore defined by the defference from 0.5.}} 
  We also define the counterparts in the whole electorate in the same way (i.e., \textit{Global Majority}, \textit{Global Majority Share}, \textit{Global Minority}, and \textit{Global Minority Share}).

  The number of voters in the human group (i.e., \textit{Local Population}) is a variable directly associated with the pivotality of each vote within the group. The larger the number of voters, the less likely each vote is to affect the outcome. The share of the human group in the whole electorate (i.e., \textit{Local Population Share}) is related to the pivotality of the human group in the result. The larger the share of the human group, the more likely each vote affects the outcome. 
		
		We have seven control variables for each subject's characteristics based on the post-experimental questionnaire. 
		Specifically, we measured the extent to which the subjects feel obliged to vote, the extent to which they feel bothersome to vote, the subject's perception of electoral effectiveness in general, and the degree of prosociality (putting the benefit of the whole ahead of the benefit of the individual). 
		Subjects are also asked about their biological gender, whether or not they are science majors, and their experience of voting in national or local elections in the past (never or at least once). 
  
  \bigskip
         \noindent
 \textbf{Overall results}

We mainly focus on the behavioral effects of majority/minority status and majority/minority share on the turnout deviation, i.e., [Experiment]$-$[Theory] 
in Columns (3) and (4) of Table \ref{tab:regression_dataonly_and_deviation_controlsomittted}.
The signs of these effects and their relative magnitudes between WTA and PR are broadly consistent with our discussion of the Titanic and behavioral bandwagon effects based on the aggregated data in the previous subsections.

\bigskip
\noindent
\textbf{The effects of minority status and minority share}

Local minority status negatively affects the turnout deviation.
In Columns (3) and (4) of Table \ref{tab:regression_dataonly_and_deviation_controlsomittted}, this can be seen from the negative significant coefficient of \textit{Local Minority} for both WTA and PR. 
On the other hand, the coefficient of \textit{Global Minority} (i.e., being a majority in society) is negative, but not significant. The difference may be due to the fact that the subjects were not explicitly informed of their global minority/majority status, while their local minority/majority status was explicitly indicated on the screen.\footnote{The decision screen showed the local support rates for the candidates in the human group, but not the global support rates in society as a whole (Figure \ref{ztuenv}). The only available information about the other two (computer) groups was their population sizes and the probabilities of their members voting for each candidate and abstaining. This information might have provided, at best, clues about the global support rates.}

The turnout deviation for minority voters decreases as the minority share decreases, at both local and global levels.
This can be observed from the positive significant coefficients of the interaction terms \textit{Local Minority$\times$Local Minority Share} and \textit{Global Minority$\times$Global Minority Share} in Columns (3) and (4),
suggesting that the degree of the Titanic effect increases as the degree of minority intensifies.
Moreover, these variables also have positive significant coefficients in Columns (1) and (2), implying that as the minority share decreases, minority turnout itself decreases, and this decrease is sharper than theory.

\bigskip
\noindent
\textbf{The effects of majority status and majority share}

Similar observations can be made for the behavior of majority voters. Local majority status positively affects the turnout deviation at least under PR. In Columns (3) and (4) of the table, this can be observed from the positive significant coefficient of \textit{Local Majority} for PR. The coefficient is positive, but not significant for WTA. The coefficient of \textit{Global Majority} is not significant, for which the same explanation as for \textit{Global Minority} may apply.

The turnout deviation for majority voters increases as the majority share increases, as can be observed from the positive significant coefficients of \textit{Local Majority$\times$Local Majority Share} and \textit{Global Majority$\times$Global Majority Share}.
The degree of the behavioral bandwagon effect increases as the degree of majority intensifies.
Moreover, \textit{Local Majority$\times$Local Majority Share} is negative significant in Columns (1) and (2), implying that as the local majority share increases, majority turnout itself decreases, and this decrease is less intensive than theory. On the other hand, \textit{Global Majority$\times$Global Majority Share} is positive significant in Columns (1) and (2), that is, majority turnout itself increases as the global majority share increases.

\bigskip
\noindent
\textbf{Comparison of rules}

How do these effects vary across the rules?
The effect of local minority status (i.e., \textit{Local Minority}) does not significantly differ between WTA and PR. However, as the local minority share decreases, the turnout deviation for minority voters decreases more substantially under PR than under WTA. Indeed, in Columns (3) and (4) of the table, the coefficient of \textit{Local Minority$\times$Local Minority Share} under PR is more than twice that under WTA. This is consistent with what we observed from the aggregated data in Section 4.1: in Panel B of Table \ref{tab:tests_TheoExpe}, the deviation of minority turnout from the theoretical prediction is larger in absolute value for PR than WTA, for most voting configurations in categories Local and Both.

The effect of local majority status (i.e., \textit{Local Majority}) is significant for PR, but not for WTA. Moreover, as the local majority share increases, the turnout deviation for majority voters increases more sharply under PR than under WTA, which can be seen by comparing the coefficients of \textit{Local Majority$\times$Local Majority Share} between PR and WTA. This is somewhat consistent with Panel A of Table \ref{tab:tests_TheoExpe}, in which the deviation of majority turnout from theory is larger for PR than WTA, at least for the voting configurations in category Local. 

Finally, the corresponding effects for global minority/majority voters do not significantly differ between the two rules. This can be checked by comparing the coefficients of \textit{Global Minority/Majority} and their interaction terms with the minority/majority share between WTA and PR. 

\bigskip

\noindent
\textbf{Other findings}

The coefficients of \textit{Local Population} on turnout, Columns (1) and (2), are negative significant (only for WTA), whereas the coefficients on the turnout deviation, Columns (3) and (4), are positive significant for both rules. This is consistent with the general wisdom that turnout decreases as the pivotal probability decreases in large election, but turnout  in the experiment did not decrease as much as theory suggests.

The coefficients of \textit{Local Population Share} on turnout, Columns (1) and (2), are positive significant for both rules. This is also consistent with the theory that turnout in a group increases as the group becomes more pivotal in society. The coefficient on the turnout deviation, Columns (3) and (4), is positive significant for WTA, and negative and significant for PR, suggesting that turnout increases more than theory under WTA, and less than theory under PR.

The variables on subjects' individual characteristics are not significant, suggesting that individual characteristics did not affect turnout behavior of the subjects in our experiment.\footnote{The only exception was that the variable ``feeling obliged to vote'' was significant at the 10 percent level in Column (1). }

	\section{Conclusion}
	We conducted an online experiment in a controlled environment to examine the turnout and welfare achieved under two groupwise vote aggregation rules. Votes are first aggregated to determine the weight allocation between candidates in each group, which is then summed to determine the winner. We observed the Titanic effect and the behavioral bandwagon effect, i.e., subjects from a minority camp were less likely to vote than theoretically predicted, while those from a majority camp were more likely. Such effects were observed more acutely under the proportional rule than under the winner-take-all rule.
	
	One of the specific features of our experiment, which differs from existing ones in the literature, is that there are multiple groups within the whole electorate. Consequently, we can distinguish the effects of being in the local majority or minority from those of being in the global majority or minority on voters' decisions to turn out. From our regression results, we observe that the turnout rate correlates more strongly with being \textit{local} majority or minority than with being \textit{global} majority or minority.
	
	The consequences of these effects on democratic decision making are twofold. The first concerns efficiency of the electoral outcome. We observe that the winning probability of the majority candidate is higher in the experiment than in equilibrium, due to the high (resp. low) turnout of the majority (resp. minority) supporters. The second concerns the distribution of welfare among voters. Expected welfare increased (resp. decreased) in the majority (resp. minority) camp, suggesting that the above-mentioned behavioral effects may lead to an undesirable outcome in terms of equality, in exchange for the gain in efficiency.
	
	In our experiment, we assigned human subjects to one of the three groups, with the other two groups consisting of automated voters adopting the equilibrium strategy. This method lightened cognitive burden on the subjects, making it easier for them to understand the overall election structure. It also enabled us to eliminate data dependence across groups and make our statistical analyses tractable. However, given that the automated voters' equilibrium play may differ depending on the aggregation rules, direct comparison of the rules through our experimental results obtained by the composition of the human subjects and automated voters has become less obvious. We leave the fully comprehensive analysis comparing our results with those obtained when all groups are made up of human subjects for future research.
	
	






	\bibliography{wtaexp}
\end{document}

%% file: Table_voterconfig.tex
	\begin{table}[t]
		\caption{Voter configurations}	\label{table:config}
		\begin{center} 
				\begin{tabular}{c|c|ccc|cccc}
                    \hline
					Configuration & Category & $n_1$ & $n_2$ & $n_3$ & $p_1$ & $p_2$ & $p_3$ & $\bar{p}$ \\
					\hline
					1 & IC & 21 & 21 & 21 & 0.5 & 0.5 & 0.5 &  0.5 \\
					2 & IC & 21 & 21 & 7 & 0.5 & 0.5 & 0.5 &  0.5 \\
					3 & IC & 21 & 21 & 3 & 0.5 & 0.5 & 0.5 &  0.5 \\
					4 & Global & 21 & 21 & 21 & 0.5 & 0.5 & 0.35 &  0.45 \\
					5 & Local & 21 & 21 & 21 & 0.1 & 0.7 & 0.7  & 0.5  \\
					6 & Both & 21 & 21 & 21 & 0.35 & 0.5 & 0.5 &  0.45 \\
					7 & Global & 21 & 21 & 7 & 0.5 & 0.5 & 0.35 &  0.48 \\
					8 & Both & 21 & 21 & 7 & 0.48 & 0.48 & 0.48 &  0.48 \\
					9 & Both & 21 & 21 & 7 & 0.45 & 0.5 & 0.5 &  0.48 \\
					10 & IC & 7 & 7 & 7 & 0.5 & 0.5 & 0.5 &  0.5 \\
					11 & IC & 7 & 21 & 21 & 0.5 & 0.5 & 0.5 &  0.5 \\
					12 & Both & 7 & 21 & 21 & 0.15 & 0.5 & 0.5 &  0.45 \\
					13 & Local & 7 & 21 & 21 & 0.1 & 0.57 & 0.57   &  0.5 \\
					14 & Global & 7 & 7 & 7 & 0.5 & 0.5 & 0.35 &  0.45 \\
					15 & Local & 7 & 7 & 7 & 0.1 & 0.7 & 0.7  & 0.5 \\
					16 & Both & 7 & 7 & 7 & 0.35 & 0.5 & 0.5 &  0.45 \\
					17 & Both & 7 & 21 & 21 & 0.48 & 0.48 & 0.48 &  0.48 \\
					18 & Global & 7 & 21 & 21 & 0.5 & 0.5 & 0.45 &  0.48 \\
					\hline
				\end{tabular}
		\end{center}
	\end{table}

%% file: table_turnoutrates.tex
	\begin{table}[]
		\centering
		\caption{Comparison of turnout rates: equilibrium and experiment.}
			\begin{tabular}{ccllllllll}
				\hline
				\multicolumn{1}{c}{\multirow{2}{*}{Config.}} & \multicolumn{1}{c}{\multirow{2}{*}{Cat.}} & \multicolumn{4}{c}{Equilibrium }         & \multicolumn{4}{c}{Experiment (average)}      \\ \cline{3-10} 
				\multicolumn{1}{c}{}     & \multicolumn{1}{c}{}  & $t_{1,A}^{\rm WTA}$ & $t_{1,B}^{\rm WTA}$ & $t_{1,A}^{\rm PR}$ & $t_{1,B}^{\rm PR}$ &$\hat{t}_{1,A}^{\rm WTA}$ & $\hat{t}_{1,B}^{\rm WTA}$ & $\hat{t}_{1,A}^{\rm PR}$ & $\hat{t}_{1,B}^{\rm PR}$ \\ \hline
				1    & IC & 0.359 & 0.359 & 0.391& 0.391 & 0.449 & 0.473 & 0.495& 0.448 \\
				2    & IC & 0.359 & 0.359 & 0.424& 0.424 & 0.471 & 0.493 & 0.504& 0.462 \\
				3    & IC & 0.359 & 0.359 & 0.442& 0.442 & 0.481 & 0.477 & 0.489& 0.467 \\
				4    & Global & 0.359 & 0.359 & 0.367& 0.383 & 0.340 & 0.486 & 0.321& 0.496 \\
				5    & Local & 0.283 & 0.144 & 0.753& 0.250 & 0.071 & 0.291 & 0.103 & 0.457\\
				6    & Both & 0.368 & 0.301 & 0.437& 0.333 & 0.184 & 0.444 & 0.218& 0.522 \\
				7    & Global & 0.359 & 0.359 & 0.417& 0.425 & 0.410 & 0.440 & 0.415& 0.491 \\
				8    & Both & 0.363 & 0.353 & 0.427& 0.416 & 0.340 & 0.465 & 0.301& 0.471 \\
				9    & Both & 0.368 & 0.345 & 0.441& 0.405 & 0.288 & 0.494 & 0.314& 0.559 \\
				10   & IC & 0.516 & 0.516 & 0.553& 0.553 & 0.420 & 0.532 & 0.485& 0.451 \\
				11   & IC & 0.516 & 0.516 & 0.421& 0.421 & 0.410 & 0.408 & 0.434& 0.375 \\
				12   & Both & 0.585 & 0.344 & 0.630& 0.301 & 0.188 & 0.407 & 0.113& 0.428 \\
				13   & Local & 0.588 & 0.307 & 0.737 & 0.287& 0.121 & 0.309 & 0.161 & 0.369\\
				14   & Global & 0.516 & 0.516 & 0.526& 0.555 & 0.329 & 0.543 & 0.306& 0.489 \\
				15   & Local & 0.538 & 0.284 & 1.000& 0.354 & 0.162 & 0.309 & 0.195 &0.455\\
				16   & Both & 0.553 & 0.455 & 0.629& 0.480 & 0.232 & 0.530 & 0.175& 0.540 \\
				17   & Both & 0.521 & 0.508 & 0.424& 0.414 & 0.322 & 0.472 & 0.285& 0.392 \\
				18   & Global & 0.516 & 0.516 & 0.415& 0.422 & 0.370 & 0.479 & 0.362& 0.367 \\ \hline
			\end{tabular} \label{tab:comparison_turnout}
	\end{table}

%% file: table7new_BandwagonTitanic.tex
\begin{table}[]

			\centering
			\caption{Difference of turnout rates, [Experiment]$-$[Theory].}

\subcaption*{Panel A: Majority - Behavioral bandwagon effect.}

\begin{tabular}{lwc{19mm}wc{19mm}wc{19mm}wc{19mm}wc{19mm}wc{19mm}}
\hline \hline
WTA    &           &         &          &        &         &        \\
Global & config. 4 & 7       & 14       & 18     &         &        \\
       & 0.127     & 0.081   & 0.027    & $-$0.037 &         &        \\
Local  & config. 5 & 13      & 15       &        &         &        \\
       & 0.146     & 0.002** & 0.025    &        &         &        \\
Both   & config. 6 & 8       & 9        & 12     & 16      & 17     \\
       & 0.144     & 0.112*  & 0.149**  & 0.062  & 0.075** & $-$0.035 \\ \hline
PR     &           &         &          &        &         &        \\
Global & config. 4 & 7       & 14       & 18     &         &        \\
       & 0.113     & 0.066** & $-$0.066   & $-$0.055 &         &        \\
Local  & config. 5 & 13      & 15       &        &         &        \\
       & 0.207***  & 0.082   & 0.101**  &        &         &        \\
Both   & config. 6 & 8       & 9        & 12     & 16      & 17     \\
       & 0.189***  & 0.055   & 0.154*** & 0.127  & 0.059** & $-$0.022 \\ 
       \hline \hline
\end{tabular}
\bigskip

\subcaption*{Panel B: Minority - Titanic effect.}
\begin{tabular}{lwc{19mm}wc{19mm}wc{19mm}wc{19mm}wc{19mm}wc{19mm}}
\hline \hline
WTA       &           &           &           &           &           &    \\
Global    & config. 4 & 7         & 14        & 18        &           &    \\
&$-$0.019    & 0.051     & $-$0.187*** & $-$0.146*** &           &               \\
Local     & config. 5 & 13        & 15        &           &           &    \\
&$-$0.212*** & $-$0.467*** & $-$0.376*** &           &           &               \\
Both      & config. 6 & 8         & 9         & 12        & 16        & 17 \\
&$-$0.183*** & $-$0.023    & $-$0.080*    & $-$0.397*** & $-$0.321*** & $-$0.200***     \\ \hline
PR        &           &           &           &           &           &    \\
Global    & config. 4 & 7         & 14        & 18        &           &    \\
&$-$0.046    & $-$0.002    & $-$0.220***  & $-$0.053*   &           &               \\
Local     & config. 5 & 13        & 15        &           &           &    \\
&$-$0.650***  & $-$0.576*** & $-$0.805*** &           &           &               \\
Both      & config. 6 & 8         & 9         & 12        & 16        & 17 \\
&$-$0.219*** & $-$0.126*** & $-$0.127**  & $-$0.516*** & $-$0.454*** & $-$0.138***     \\ 
\hline \hline
					\addlinespace[1ex]
					\multicolumn{5}{l}{\footnotesize \textsuperscript{***}$p<0.01$, 
						\textsuperscript{**}$p<0.05$, 
						\textsuperscript{*}$p<0.1$,  Wilcoxon one-sample test.} 
\end{tabular}
			\label{tab:tests_TheoExpe}
\end{table}

%% file: table_WelfareComparison.tex
\caption{Average welfare comparison, theory vs. experiment.}
\centering
\begin{tabular}{lccclccc}
\hline
\multicolumn{4}{l}{Expected welfare of majority}                                                      &                      &                      &                      &                      \\
         & \multicolumn{3}{c}{WTA}                                               &  & \multicolumn{3}{c}{PR}                                               \\ \cline{2-4} \cline{6-8} 
         & Theory                  & Experiment           & Diff.(\%)      &  & Theory                 & Experiment           & Diff.(\%)      \\ \hline
Global   & 0.552                  & 0.598               & +8.3\%              &  & 0.563                 & 0.610               & +8.4\%              \\
Local    & 0.442                  & 0.466               & +5.4\%              &  & 0.553                 & 0.709               & +28.1\%             \\
Both     & 0.560                  & 0.641                & +14.4\%             &  & 0.571                 & 0.712               & +24.6\%             \\ \hline
\multicolumn{4}{l}{Expected welfare of minority}                                                      &                      &                      &                      &                      \\
         & \multicolumn{3}{c}{WTA}                                               &  & \multicolumn{3}{c}{PR}                                               \\ \cline{2-4} \cline{6-8} 
         & Theory                  & Experiment           & Diff.(\%)      &  & Theory                 & Experiment           & Diff.(\%)      \\ \hline
Global   & 0.409                  & 0.343                & $-16.1\%$             &  & 0.398                 & 0.336               & $-15.6\%$             \\
Local    & 0.527                  & 0.507               & $-3.8\%$              &  & 0.367                 & 0.252               & $-31.4\%$             \\
Both     & 0.403                  & 0.310               & $-23.0\%$             &  & 0.387                 & 0.239               & $-38.3\%$             \\ 
\hline
\multicolumn{4}{l}{Ex ante Gini coefficient}                                                      &                      &                      &                      &                      \\
         & \multicolumn{3}{c}{WTA}                                            &                      & \multicolumn{3}{c}{PR}                                             \\ \cline{2-4} \cline{6-8} 
         & Theory               & Experiment           & Diff.(\%)      &                      & Theory               & Experiment           & Diff.(\%)      \\ \hline
Global   & 0.074               & 0.135               & +81.7\%             & \multicolumn{1}{c}{} & 0.086               & 0.145               & +69.0\%             \\
Local    & 0.040               & 0.034               & $-15.2$\%             & \multicolumn{1}{c}{} & 0.031               & 0.060               & +92.6\%             \\
Both     & 0.060               & 0.137               & +128.2\%            & \multicolumn{1}{c}{} & 0.073               & 0.193               & +163.6\%             \\ \hline
\end{tabular}